\documentclass[sigconf,nonacm]{acmart}

\newcommand{\algo}{\textbf{\textsf{EneRex}}} 
\newcommand{\dataset}{\textbf{\textsf{CORLL}}} 

\renewcommand\footnotetextcopyrightpermission[1]{} % removes footnote with conference information in first column
\AtBeginDocument{%
  \providecommand\BibTeX{{%
    \normalfont B\kern-0.5em{\scshape i\kern-0.25em b}\kern-0.8em\TeX}}}

\setcopyright{acmcopyright}
\copyrightyear{2022}
\acmYear{2022}
\acmDOI{10.1145/1122445.1122456}
\acmConference[KDD '22]{KDD '22: ACM SIGKDD International conference on Knowledge Discovery and Data mining (KDD)}{August 14--18, 2022}{Washington, DC, USA}
\acmBooktitle{KDD '22: SIGKDD International conference on Knowledge Discovery and Data mining, August 14--18, 2022, Washington, DC, USA}
\acmPrice{15.00}
\acmISBN{978-1-4503-XXXX-X/18/06}

\usepackage[colorinlistoftodos]{todonotes}
\usepackage{multirow}
\usepackage{subcaption}
\usepackage{graphicx}
\usepackage{booktabs}
\usepackage{hhline}
\usepackage{dblfloatfix}
\begin{document}

\title[Lessons from {Deep Learning} applied to {Scholarly Information} Extraction]{Lessons from {Deep Learning} applied to {Scholarly Information} Extraction: What Works, What Doesn’t, and Future Directions}

\iffalse

% new author code
\settopmatter{authorsperrow=1} %make the template consider one author per row
\newcommand{\tsc}[1]{\textsuperscript{#1}} %shorthand for superscripts
\author{Raquib Bin Yousuf,\tsc{1}{${^*}$} Subhodip Biswas\tsc{1}}
\authornote{Author names arranged in alphabetical order of their first names.}
\author{Kulendra Kumar Kaushal\tsc{2}, Sathappan Muthiah\tsc{3}}
\authornote{The work was done when the authors were at Virginia Tech.}

\author{James Dunham\tsc{4}, Rebecca Gelles\tsc{4}, Nathan Self\tsc{1}, Patrick Butler\tsc{1}, Naren Ramakrishnan\tsc{1}} %to break line, start another author block

\affiliation{
%   \institution{\vskip .2cm}%add some spacing if needed
  \institution{1. Virginia Tech, Arlington, VA}
  \institution{2. Bloomberg, New York City, New York}
  \institution{3. eBay, San Jose, California}
  \institution{4. Georgetown University, Washington, D.C.}
}

\fi

% \iffalse

\author{Raquib Bin Yousuf }
\authornote{Equal contribution. Author names arranged in reverse alphabetical order.}
\author{Subhodip Biswas}
    \orcid{0000-0002-7065-2966}
    \authornotemark[1]
 \affiliation{%
  \department{Computer Science}
  \institution{Virginia Tech}
  \city{Arlington}
  \state{VA}
  \country{USA}}
 \email{raquib@vt.edu}
 \email{subhodip@vt.edu}

  \author{Kulendra Kumar Kaushal} \authornotemark[2]
 \affiliation{%
 %\department{Center for Security and Emerging Technology}
  \institution{Bloomberg}
  \city{New York City}
  \state{New York}
  \country{USA}}
 \email{kulendra@vt.edu}

 \author{James Dunham}
  \author{Rebecca Gelles}
 \affiliation{%
 \department{CSET}
  \institution{Georgetown University}
  \city{Washington, D.C.}
  \country{USA}}
 \email{james.dunham@georgetown.edu}
 \email{rebecca.gelles@georgetown.edu}

 \author{Sathappan Muthiah}\authornote{The work was done when the author was at Virginia Tech.}
 \affiliation{%
 %\department{Center for Security and Emerging Technology}
  \institution{eBay}
  \city{San Jose}
  \state{California}
  \country{USA}}
 \email{sathap1@vt.edu}

  \author{Nathan Self}
  \author{Patrick Butler}
  \author{Naren Ramakrishnan}
 \affiliation{%
  \department{Computer Science}
  \institution{Virginia Tech}
  \city{Arlington, VA}
  \country{USA}}
 \email{nwself@vt.edu}
 \email{pabutler@vt.edu}
 \email{naren@cs.vt.edu}
% % %%
% \fi

\renewcommand{\shortauthors}{Yousuf and Biswas, et al.}

%% omitted for CSET report
\begin{abstract}
Understanding key insights from full-text scholarly articles is essential as it enables us to determine interesting trends, give insight into the research and development, and build knowledge graphs. However, some of the interesting key insights are only available when considering full-text. Although researchers have made significant progress in information extraction from short documents, extraction of scientific entities from full-text scholarly literature remains a challenging problem. This work presents an automated \textbf{En}d-to-end \textbf{R}esearch Entity \textbf{Ex}tractor called\textbf{~\algo~}to extract technical facets such as dataset usage, objective task, method from full-text scholarly research articles. Additionally, we extracted three novel facets, e.g., links to source code, computing resources, programming language/libraries from full-text articles.
We demonstrate how{~\algo~}is able to extract key insights and trends from a large-scale dataset in the domain of computer science. We further test our pipeline on multiple datasets and found that the \algo~improves upon a state of the art model. We highlight how the existing datasets are limited in their capacity and how \algo~may fit into an existing knowledge graph. We also present a detailed discussion with pointers for future research. Our code and data are publicly available at \textcolor{blue}{ \href{https://github.com/DiscoveryAnalyticsCenter/EneRex}{https://github.com/DiscoveryAnalyticsCenter/EneRex}
}.
\end{abstract}






\maketitle

\section{Introduction}
% Need complete overhaul
% Objective

% 0171bdeb1c6e333287be655c667cfba5edb89b76, this was old
% 1611.08036 new one \cite{kumra2017robotic}
% General Motivation, the number of published articles and the need to extract entities:

% There are two aspects here, we are extracting from full-text and using that on large scale dataset to find out trends, check abstract

The number of scientific scholarly articles published each year is staggeringly high and continues to rise. According to DBLP, $400$k articles were published in Computer science(CS)-based research areas in 2020 alone. Extracting key scientific insights from these papers is imperative for understanding emerging technologies, their prevalence, and relationships, and for enabling analysts and policymakers to identify key trends. Information extraction of these entities from large-scale datasets would facilitate the creation of structured knowledge graphs. Existing work builds these knowledge graphs from citation graphs and clusters, coupled with the classification of the papers by various conferences or libraries, such as the CSET Map of Science%\footnote{\textcolor{blue}{https://sciencemap.cset.tech}}
\cite{aI_funding_research, cset_Dunham_ai} and the Microsoft Academic Graph \cite{mag_mas}. These knowledge graphs can be used to discover clusters of papers belonging to topics of research. Recently, researchers have been attempting to automatically classify documents~\cite{cohan2020specter} and discover clusters of paper related to a specific task~\cite{dunham_ai_trends}. Moreover, such knowledge graphs are already in use helping policy makers look into research funding practices in artificial intelligence~\cite{aI_funding_research}. We believe that actual scientific entities from the papers would complement the existing citation-based knowledge graphs with more in-depth knowledge and further enrich the information. Motivated by this, we propose an information extraction pipeline for extracting technical entities from the full text of research articles, allowing us to establish trends present in large scholarly databases, particularly in the domain of CS.

% We focus on the extraction of source code link, dataset usage, objective task, method, computational platform, employed hardware resources, computation time of the algorithm, and its language/library dependencies.
% Each of the entities in question has huge interest from the research community and policymakers. Source code links would be helpful to follow the actual projects of researchers. Identifying objective tasks and methods would enable us to classify the articles by their research areas and methodologies. Various Machine Learning (ML) or Deep Learning (DL) algorithms need different compute platforms to run and produce the desired results. Extracting these platforms from a wide range of research articles gives us a mapping between DL algorithms and their computational needs. We conducted some experiments to demonstrate how the developed system can be helpful to establish trends.

% \usepackage{booktabs}
\begin{table*}[bp!]
\caption{Comparison of \algo~with similar work}
\label{tab:similar_work}
\begin{tabular}{@{}|l|c|c|c|c|c|c|c|c|c|@{}}
\hline
Method & \multicolumn{1}{l|}{\begin{tabular}[c]{@{}l@{}}Extract\\ Dataset\end{tabular}} & \multicolumn{1}{l|}{\begin{tabular}[c]{@{}l@{}}Extract \\ Objective \\ Task\end{tabular}} & \multicolumn{1}{l|}{\begin{tabular}[c]{@{}l@{}}Extract\\ Method\end{tabular}} & \multicolumn{1}{l|}{\begin{tabular}[c]{@{}l@{}}Full Text\\ Extraction\end{tabular}} & \multicolumn{1}{l|}{\begin{tabular}[c]{@{}l@{}}Extract\\ Computing \\ Resources\end{tabular}} & \multicolumn{1}{l|}{\begin{tabular}[c]{@{}l@{}}Extract\\ Language\\ Library\end{tabular}} & \multicolumn{1}{l|}{\begin{tabular}[c]{@{}l@{}}Extract\\ Links to\\ Source Code\end{tabular}} & \multicolumn{1}{l|}{\begin{tabular}[c]{@{}l@{}}No Need for\\ Annotation\end{tabular}} & \multicolumn{1}{l|}{\begin{tabular}[c]{@{}l@{}}Simplified\\ Pipeline\end{tabular}} \\ \hline
% DyGIE++\cite{Wadden2019EntityRADygie++} & \checkmark  & \checkmark  & \checkmark  &  &  &  &  &  &  \\ \hline
SciREX\cite{scirex} & \checkmark  & \checkmark  & \checkmark  & \checkmark  &  &  &  &  &  \\ \hline
\algo & \checkmark  & \checkmark  & \checkmark  & \checkmark  & \checkmark  & \checkmark  & \checkmark  & \begin{tabular}[c]{@{}c@{}}partially\end{tabular} & \checkmark  \\ \hline
\end{tabular}
\end{table*}

% challenges:
Identifying salient scientific facets, or entity types, and extracting them from scholarly articles is an important research endeavor in the information retrieval community~\cite{gupta2011analyzing, luan2018information}. Extraction involves encoding texts, identifying sections where relevant information is present, and then extracting the required information in a structured format. Most research work in this area has traditionally involved working with machine-readable metadata such as title, abstract, etc.~\cite{tateisi-etal-2016-typed, lakhanpal2015towards}. 
%cite needed, manning and other, look for new ones as well, tateisi, lakhanpal, semeval
However, many important facets can only be extracted when the full text is taken into consideration since such information is not available in the metadata alone.
A recent survey paper concluded that the majority of work on key insight extraction uses abstracts only~\cite{nasar2018information}. They also concluded that the primary challenge of full-text analytics is that the complexity of the manual annotation processes grows as the dataset grows. 
% Usually, the length of a research paper is between eight to fourteen pages, and extraction involves processing long documents. 
Identifying which section of an article contains information relevant to a given entity type is challenging since much of the text data is irrelevant to that particular entity type. Previous work has focused primarily on information extraction from shorter documents like news articles, blogs, user posts, and comments on different social media platforms \cite{ritter2011named,brin1998extracting,arulanandam2014extracting}. 
Instead, we focused on developing a full-text research entity extraction system which will enable us to discover trends in computer science research.

The scarcity of large-scale data to train and evaluate our models remains a challenge. Many publicly available, labeled datasets contain annotations for small documents. However, there is a lack of ground truth data for full-text scientific articles. We tackled this issue with an automated data generation process based on syntactic patterns to generate training data.
Recently, there have been attempts to extract information from full-text scholarly articles and create full-text datasets that can help train and build full-text extraction modules, for instance, see~\cite{scirex}. However, existing datasets are limited in their capacity. In some cases, ground truth data does not have all the entities used for every paper. Moreover, to the best of our knowledge, three of our facets (links to source code, computing resources, and language/library used) have not been extracted in prior work, so no ground truth datasets are available for those facets.

\begin{figure}
  \centering
  \includegraphics[width=\linewidth]{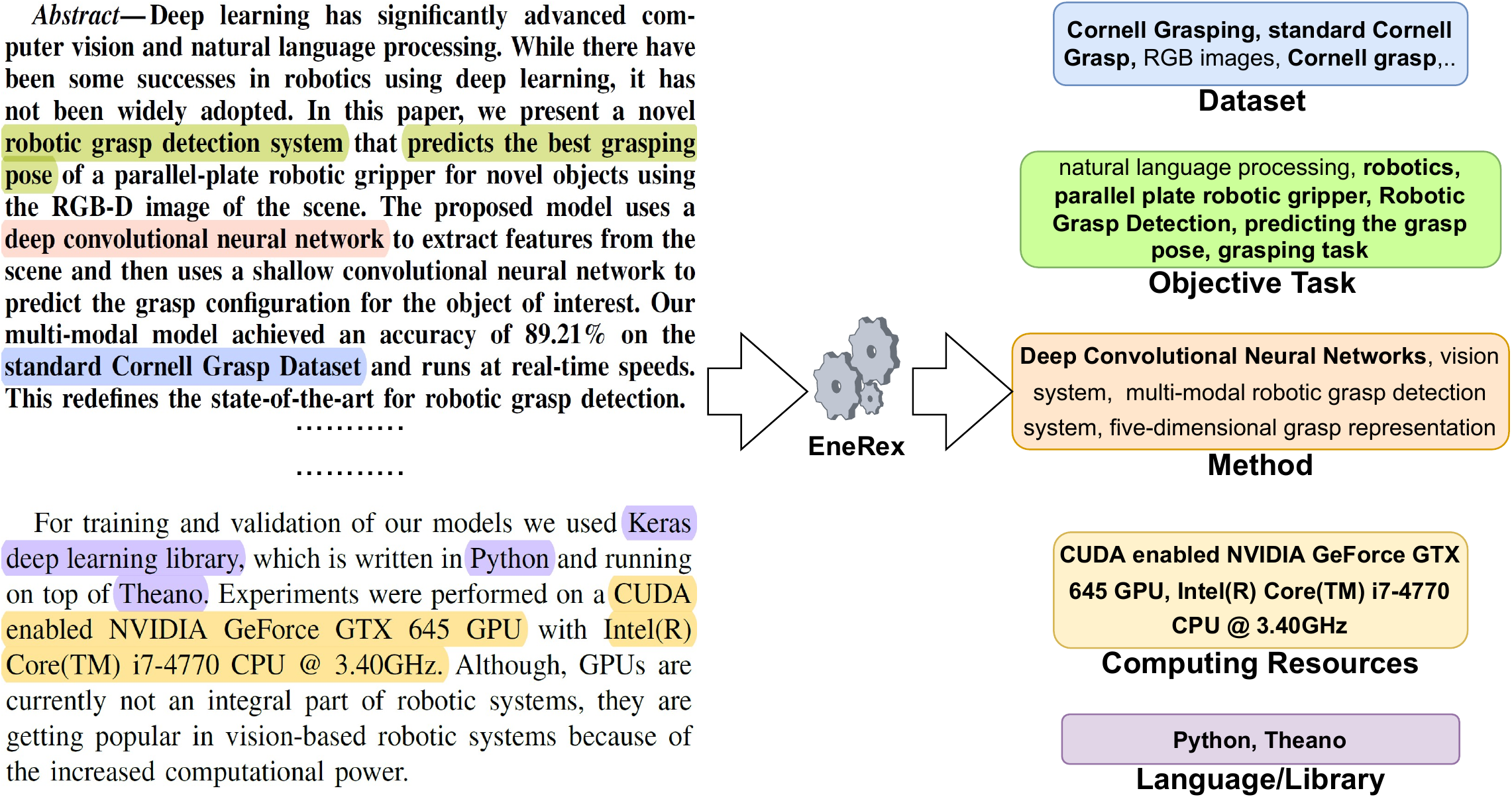}
  \caption{\algo~ in practice.}
  \label{fig:EneRex_system_in_practice}
\end{figure}

%intro of our model:
In this work, we introduced a DL-based automated tool named End-to-end Research Entity Extractor (\algo) to extract from a paper's full text six entity types: links to source code, names of any datasets used, the paper's objective task, the method by which the paper attempts the object task, how many computing resources were required, which programming languages were used, and which programming libraries were used. We ran \algo~on a CS-based large-scale scholarly dataset and used the extracted information to discover trends and insights about computer science research areas. We discussed how \algo~can fit into an existing knowledge graph to complement that information. We also came up with a novel dataset for computing resources and language/library (CORLL). 
\algo~uses transfer learning to generate scientific contextual embeddings from SciBERT~\cite{beltagy2019scibert}, which enables it to learn a better representation of scholarly articles. SciBERT extends the BERT model~\cite{devlin2018bert} and has been trained on scientific texts. \algo~was able to improve upon SciREX~\cite{scirex} which is the only model aiming to extract similar facets from full-text scholarly articles, to the best of our knowledge. In addition to the facets provided by SciREX, we also extracted computing resources and any languages or libraries used in a paper. \algo~is computationally simple with different components for handling salience and entities. For generating training data, \algo~does not need manual annotations for all of the facets, and SciREX requires more involved annotation efforts for all the facets. Table \ref{tab:similar_work} shows a comparison between existing models and \algo. An example of \algo~in practice is shown in Figure \ref{fig:EneRex_system_in_practice}.

The remainder of the paper is organized as follows. Section~\ref{sec:scientific} discusses the scientific entities in research articles and previous works to identify those. We introduce the \algo~system in Section~\ref{sec:system} followed by evaluation in Section~\ref{sec:empirical}. Our system is used to discover trends in computer science scholarly articles in Section~\ref{sec:experiment}. Finally, in Section~\ref{sec:discussion} we present some observations and conclude the paper in Section~\ref{sec:conclusion}.

\section{Scientific Facets in Research Articles and Related Works}
\label{sec:scientific}
% introduce which are some important sci enti in research articles. Clearly discuss why did we chose to extract those. 
% Also add some literature review, working on these entities
% Does the following belong here? These are discussion for need to research hardware resources:
The traditional facets extracted by the information retrieval community from scientific papers have been objective task, method, and material used. These three facets are known to be helpful to researchers for quickly comprehending the main thrust of a scientific article. Gupta et al.~\cite{gupta2011analyzing} attempted to discover the focus, technique and domain of research articles via pattern matching in a bootstrapping manner. Tsai et al.~\cite{tsai2013concept} also used bootstrapping to identify two categories of concepts: techniques and application. 

More recent approaches consist of applying neural models to identify entities and their relations. Luan~\cite{luan2018information} used neural models with semi-supervised learning to extract entities and their relationships. Mesbah et al.~\cite{Mesbah2018TSENERAI} developed NER models to detect scientific datasets and methods in an iterative manner. These automatic scientific entity extraction tasks revolved around using only the abstracts of scholarly articles ~\cite{augenstein-etal-2017-semeval}. SciERC~\cite{luan-etal-2018-multi} and Dygie++~\cite{Wadden2019EntityRADygie++} are two such models which work on a dataset of 500 annotated abstracts. Nasar et al.~\cite{nasar2018information} details  past research which extracted facets such as problems addressed, domain, tools, and evaluation measures. The SciREX~\cite{scirex} model works with full text documents in an end-to-end fashion, in comparison to previous research approaches. Our model adds three more features and improves upon the results from SciREX.

In an effort to improve reproducibility, researchers have increasingly been publishing their code along with their papers.
Papers with Code\cite{PwC} collects state-of-the-art papers for some research categories that have been published with source code.
The structured facets available from this website are dataset, metric, task and method.
% We decided to add source links to our intended list of facets for extraction which has never been extracted by anyone before.

% \begin{figure}[htbp!]
%   \centering
%   \includegraphics[height=6cm, width=0.9\linewidth,keepaspectratio]{Images/NER_diagram.pdf}
%   \caption{The NER architecture of \algo}
%   \label{fig:ner_archi}
% \end{figure}

Researchers have shown that complex natural language processing models like GPT~\cite{radford2018improving}, BERT~\cite{devlin2018bert}, Turing-NLG~\cite{rajbhandari2020zero} with billions of parameters are generally more efficient at solving challenging problems like information extraction, machine translation, and natural language generation. Of course,  the power consumed and the required computing resources in training such complex models is an ever-present issue~\cite{stochasticparrots}. 
% Moreover, running neural networks and other machine learning algorithms require powerful computing resources, i.e., GPU, CPU and physical memories.
In this work, we aim to extract different computing resource entities which could help us track the efficiency of different computing resources across different neural networks. Furthermore, several libraries, such as PyTorch, TensorFlow, and Caffee are available for different programming languages like Python, Java, and MATLAB. Tracking the usage of different languages and library dependencies across research articles could help researchers determine the best possible combination and help analyze usage trends. All of the six facets extracted in our system are listed in Table~\ref{tab:entity_list}.

\begin{table}[htp!]
\caption{List of facets for extraction}
\vspace{-1em}
\label{tab:entity_list}
\begin{tabular}{|l|}
\hline
\textbf{Facets}                            \\ \hline \hline
Source Code Links                              \\ \hline
Dataset used                                   \\ \hline
Objective Task                            \\ \hline
Method used                               \\ \hline
Computing Resources \\ \hline
Language/library                       \\ \hline
\end{tabular}
\vspace{-1em}
\end{table}

%% Section is omitted for CSET report
% \section{Related Work}
% % we definitely need a related work here, without it, the paper would not look good
% The information extraction from scholarly articles involves the identification of important concept, idea and techniques used in the scholarly articles. The definition and scope of the concepts as scientific entities have been evolved. 

% Any graph showing the trends? Simple would work fine. Just to show the usefulness of the system
% insight should have graphs with the data. e.g., pytorch is used more than tensorflow
% energy usage of neural network model- paper
% fairness related to method vs fairness related to dataset

% (eg - see Section 9 of https://people.cs.vt.edu/naren/papers/kddindg1572-ramakrishnan.pdf - begin with questions that the experimental results are planning to address and then address them.) 

% You probably want to show some interesting results, ie., not specific extractions but trends. 
% e.g, what is the growth of terms like NVIDIA, Deep learning, GANs, etc. 
% can you show when they started picking up? Have they peaked? etc. 
% Did the rise of term X coincide with the rise of term Y? What does that say about the field?

\section{The \algo~system}
\label{sec:system}
% system archi may have some processing information, like ingest processing etc but less is better, Prioriotize the characterization and experimentation

\begin{figure*}
  \centering
  \includegraphics[trim={0cm 0.80cm 3.80cm 0},clip, width=.99\linewidth, keepaspectratio]{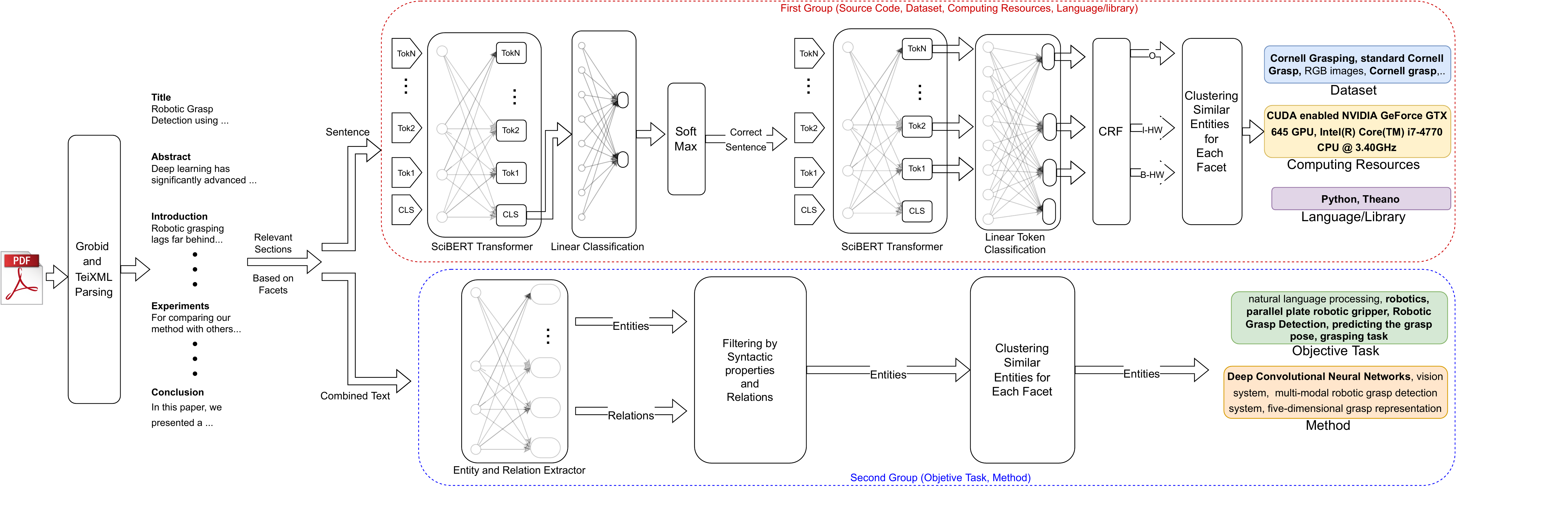}
  \caption{The architecture of the \algo~system.}
  \label{fig:system_archi}
\end{figure*}

\algo~is capable of identifying and extracting six facets from a full-text scholarly article. Extraction for these six facets is grouped into two different pipelines in terms of the adopted methodology, depending on context and other syntactic properties. The first group (source code link, dataset, computing resource and language/library) was extracted by a weakly-supervised learning task. Here, automated labeling of sentences and entities created a noisy ground truth set. For this group, \algo~ identifies entities in two steps: i) identifying the relevant sentences; ii) identifying the entities from the selected sentences. The second group (objective task and method) was extracted with the help of transfer learning. Each of the facets were designed with syntactic properties in mind. The training and prediction modules were kept separate for simplicity, thereby adopting batch training as the default knowledge ingestion process. However, all of the facet extraction tasks share a full-text extraction and ingestion module.
In order to build a structured representation from an article’s full text, \algo~can extract full-text from PDF, plain text, and JSONL.
% \algo~takes advantage of each article’s full text to build a structured representation. This required extracting full-text from article PDF. Moreover, the ingestion module is capable of handling multiple types of raw inputs, such as plain text and JSONL. 
Subsequent entity extraction sub-modules use this structured representation of the article to drive training and inference. The full-text extraction module is described first, followed by the details of each facet and module. A conceptional overview of \algo~architecture in inference is shown on Figure~\ref{fig:system_archi}.
% We describe each level in turn.

% PDF challenges, not significant to discuss in the paper but if it is, compare with S2ORC, quali/table/survey17: Sir Tim Berners-Lee [3] classifies PDF at the lowest level of data openness in his five-star deployment scheme for open data. Also, the format might not comply with accessible and machine processable open data principles, because not everyone knows how to extract data from a PDF. Hence the processing becomes more costly and challenging as there is insufficient information regarding the structure of the data contained in it [35].

% there are some citation for GRobid
\subsection{Data Ingestion}
\algo~ingests PDFs of scholarly articles. To build a large dataset of PDF files, we collected PDF files and metadata from arXiv\cite{arxiv}, an open-access research article distribution service. The arXiv corpus contains about 2 million scholarly articles in many fields. Since research articles published in fields such as physics or astronomy are not relevant to our entity extraction tasks, we collect only articles published in fields related to Computer Science.
% such as Machine Learning (LG), Hardware Architecture (AR), Computer Vision (CV), Artificial Intelligence (AI), and Information Extraction (IR). 
Using publication metadata and ``cs.'' tags from the arXiv dataset to filter out documents from other disciplines,  we collected PDFs and metadata for 241,646 articles. The full-text and metadata from each document were ingested using Grobid~\cite{GROBID} reference parsing tools which can extract, among other things, headers, references, citation contexts, and authors from article text. Tkaczyk et al.~\cite{tkaczyk2018machine} evaluated various reference parsing tools. Among those that handle full article text, Grobid was the best performing, followed by Cermine~\cite{cermine}. 
% Multiple organizations (ResearchGate, Mendeley, Internet Archive) have Grobid deployed in their production environments. 
Grobid successfully processed 240,051 documents, about 99\% and we generated structured representations with metadata, sections, bibliography entries, and footnotes for all papers.
% To build a citation graph within our dataset, we linked bibliography entries to document titles and authors using Levenshtein distance as the similarity metric for titles and authors' names.

\subsection{Training Data Generation}
For the task of extracting these facets from scholarly articles, the main challenge is the lack of ground truth available for training inference models. We tackled this issue by using weakly supervised learning tasks. For the first group of facets (source code link, dataset, computing resource and language/library), we extracted salient ground truth by following the lexical and syntactic properties of the sentences. These facets required annotations on different levels to build classifiers and named entity recognizers (NERs). The patterns were designed by taking advantage of sentence-level dependency parsing and the part-of-speech (POS) tags on each word. The goal of this automated extraction pipeline is to require the least human effort to extract each facet. However, the quality of this automated pattern based extraction varied by facets. For the source code and dataset facets, this pattern-based extraction produced less noise than computing resources and language/library, for which the pipeline produced significant noise in the entity level annotation. After identifying candidate sentences, a follow-up manual annotation process was carried out to identify the entities within the sentences for these two facets.

\paragraph{Facet 1: Source Code Links}
To extract references to source code in a paper, our data generation script first selects any sentences with references or footnotes that contain URLs. Using spaCy~\cite{spacy} for dependency parsing obtains universal dependency relation tags for the selected sentences and selects a contextual location of a term (i.e., subject, object, root, etc.). Any sentence with at least two patterns and three occurrences is considered a candidate sentence. The sample patterns are shown in Appendix \ref{app:patterns} . Special importance was given to adjective and object patterns. Sentences containing certain common words such as figure and table were excluded. These constraints were decided by empirical analysis of the results. This algorithm created a set of sentences that were used to train a sentence classifier for this facet.

\paragraph{Facet 2: Dataset Used}
This process is similar to the source code process except the number of patterns and their templates are different. There are ten patterns for extracting mentions of datasets used in a paper; seven of which depend on dependency relation tags, as shown in Appendix \ref{app:patterns}. The rest are based on whether a sentence contains a reference, footnote, URL, or well-known dataset name. As a first step, the pipeline extracts sentences containing words about using dataset materials (e.g., dataset, corpus, database etc.) and checks the next five sentences against our patterns. In practice, sentences that contain the word ``dataset'' often do not contain details of the dataset name or its usage, reserving the actual mention for subsequent sentences. We identified dataset entities by following several heuristic rules: a dataset name must be a noun or noun phrase, must start with capital, but may end with digits. We used the shortest dependency path to assign scores to each candidate entity and empirically decided on the threshold for selecting a candidate sentence and candidate entity.  These were then used to train a sentence classifier and NER for this facet.

\paragraph{Facet 5 and 6: Computing Resources, Programming Language/Library}
For these facets, our data generation script selects sentences which contain certain seed words from a set we curated. Then, Natural Language Toolkit (NLTK) is used to tokenize, lemmatize, and remove stop words from each sentence. The algorithm extracts patterns using the part-of-speech (POS) tags and handcrafted rules to identify new candidate seed words based on these patterns. Each candidate seed word is assigned  a score. If the score is over a certain threshold, the candidate seed word is appended to the original set of seed words, otherwise it is discarded. After identifying the candidate sentences, we tried an automated extraction of the entities for these two facets. However, the results were not satisfactory so we carried out a manual annotation of the entities within the sentences for training the named entity recognition (NER) model for these two facets. In this process, we tagged 600 sentences with both computing resource and language/library entities.

\paragraph{\dataset: A Novel Dataset with Computing Resources and Language/Library Entities}
As there is no previous work on extracting hardware-related entities, we propose a novel NER dataset for computing resources and language/library. \dataset is comprised of sentences containing entities that have been used and not just mentioned in the research article, and the text span in which the entity is present in the BILUO format. The dataset contains over 600 such salient sentences and around 1400 annotated entities overall. Furthermore, the computing resources facet is divided into computing platform, compute time, and hardware resources entities. Table \ref{tab:corll} shows the distribution of different facets in the \dataset~dataset, from which we can discern that some facets, like compute time, are sparse and therefore may be challenging to extract.
% The details on the annotation process and followed standards are discussed in Appendix \ref{app:appendix_one}. 

\begin{table}[htpt!]
\centering
\caption{Facets and their Distribution in CORLL}
\vspace{-1em}
\begin{tabular}{|c|c|}
\hline
\textbf{Facets} & \textbf{Count} \\ \hline
Compute Platform & 181 \\ \hline
Compute Time & 51 \\ \hline
Hardware Resources & 576 \\ \hline
Programming Language & 367 \\ \hline
Programming Library & 168 \\ \hline
\end{tabular}%
\label{tab:corll}
\end{table}

\subsection{Model Training}
For each of these four facets (source code links, dataset, computing resources, language/library), we trained separate models to identify related sentences and entities. In general, an input sentence is first passed through a SciBERT transformer model to create contextual embeddings. Then a linear sequence classification layer is trained on top of that to identify it as a ``correct'' or ``incorrect'' sentence for the respective facet. The second step is to train a NER model. For source code, the correct sentences are sufficient. For the dataset, computing resources, language/library facets, we trained three named entity recognizers. The ground truth data for each entity are first tokenized using spaCy. The tokens are tagged in BILUO format (beginning, inside, last, unit, outside) and trained using a BERT + CRF based model. The model layers are initialized with the SciBERT model, followed by token classification layer and a CRF tagger. 

During prediction, we take each sentence of a paper and feed it through the sentence classifier first. The selected sentences are then fed through the NER to get the final entity list. For datasets, there can be multiple mentions of the entities in a paper so a single dataset name can come up through multiple sentences in the system. So we further clustered similar entities for each scholarly article. For source code links, instead of an NER module, correct sentences go through a URL extraction algorithm involving the sentence and any footnotes the sentence references.

\subsubsection{Objective Task and Method}
For the other main group of facets (objective task and method), candidate entities were extracted using a state-of-the-art NER model and salience was achieved by pruning the entities using syntactic properties. In particular, \algo~took advantage of the structured full-text representation to focus this process on only the introduction, conclusion, and similar sections. 
%as input data for extracting these two facets. 
This allows more salient entities and relations to be extracted, thereby increasing the chances of finding a mentioned task or method in a document. We observed that a task and method pair often appear in a sentence connected by a `USED-FOR' relation. A task (and/or method) may be connected to the method (and/or task) through other general or specific terms (entity or not) by `PART-OF', `FEATURE-OF', or `HYPONYM-OF' relations. The entities and the relations are extracted by adopting a scientific entity and relation extractor described by Wadden et al.~\cite{Wadden2019EntityRADygie++}. We followed these heuristics and an exhaustive search algorithm with our above hypothesis to find the most salient objective task and method for an  article. \algo~clustered similar entities using fuzzy matching to get final task and method extractions for an article.

\section{Empirical Evaluation}
\label{sec:empirical}
We evaluate each of our facet extraction tasks separately against multiple gold datasets. Not all available gold datasets are suitable for our purposes since they are often designed to the needs of the researcher and community that generated them. Moreover, we could not find one encompassing all the facets we are looking for. E.g., SciREX dataset contains only three of our targeted facets. So, we extracted the appropriate facets from each dataset and matched our results against them. For computing resources and language/library dependencies, there is no established ground truth dataset available. Moreover, available datasets are limited in their capacity. We built in-house annotated ground truth sets to address such issues. The following subsections detail the results of component-based evaluation for some of the facets. We also compared \algo~with end-to-end predictions from available state-of-the-art extraction models.

\subsection{Evaluation Metrics}
\label{sec:eval_metric}
For each facet, we calculated precision, recall, and macro F1 for every facet available in a particular dataset.For links to source code, if the extracted URL matched with ground truth URL the extraction is correct. Example of cases for links to source code is given in Appendix~\ref{app:eval}
For the dataset, objective task, and method facets, we compare the gold truth with our extracted entity clusters. We used fuzzy string matching with a threshold value of 0.85, empirically determined, to decide if the gold truth entity matched with any element of the clusters. We calculated recall as how many of the gold truth were extracted by the model and precision as correct extractions divided by total number of clusters.

\subsection{Human Annotated Dataset}
We annotated 145 artificial intelligence(AI) papers from 2014-2018, roughly 30 from each year with BRAT. We tested the extractions of source code link, dataset usage, computing resources and language/library against this ground truth dataset. The distribution of different facets in the dataset are given in Appendix~\ref{app:dataset_statistics}. For most papers, the first, or most prominent, related sentence was noted because sometimes there are too many sentences that mention an entity. On the contrary, the pipeline may extract any sentence that mentions an entity, which is often more than one. The metric values for all facets are presented in Table~\ref{tab:evaluation}. A possible reason for low precision and recall on the language/library facet is that many entities in this set have too few characters, for instance the programming languages C and R which can be hard to differentiate from abbreviations and sentences with equations.

% For source code links, there were only 10 papers that mentioned any source code  used in the development of the paper. Among these, \algo~was able to extract at least one URL for 4 papers (61.83\%). Two of those four fully matched their corresponding ground truth URL and the other two partially matched. The number of correct extractions is 4. The recall by paper is 40\% and precision by paper is 100\%. Actual precision by number of extracted URLs is 28.75\%. 

% For dataset usage, 73 papers had annotated dataset entities. Our recall was 77\% and precision is 58\%. 

% \iffalse

\begin{table*}[t]
\caption{Evaluation of \algo~ on different datasets}
\vspace{-1em}
\label{tab:evaluation}
\begin{tabular}{@{}lcccccccc@{}}
\toprule
\multirow{2}{*}{\textbf{Facet}} & \multicolumn{2}{c}{\textbf{\begin{tabular}[c]{@{}c@{}}EneRex on\\ Annotated Dataset\end{tabular}}} & \multicolumn{2}{c}{\textbf{\begin{tabular}[c]{@{}c@{}}EneRex on \\ Papers with Code Dataset\end{tabular}}} & \multicolumn{2}{c}{\textbf{\begin{tabular}[c]{@{}c@{}}EneRex on\\ SciREX Dataset\end{tabular}}} & \multicolumn{2}{c}{\textbf{\begin{tabular}[c]{@{}c@{}}SciREX on SciREX Dataset\\ Salient Entity Clusters\\ (By our evaluation method)\end{tabular}}} \\  \cline{2-9}  %\cmidrule(l){2-9} 
 & \multicolumn{1}{|c|}{\textbf{Precision}} & \multicolumn{1}{c|}{\textbf{Recall}} & \multicolumn{1}{c|}{\textbf{Precision}} & \multicolumn{1}{c|}{\textbf{Recall}} & \multicolumn{1}{c|}{\textbf{Precision}} & \multicolumn{1}{c|}{\textbf{Recall}} & \multicolumn{1}{c|}{\textbf{Precision}} & \multicolumn{1}{c|}{\textbf{Recall}} \\ \hline
\multicolumn{1}{|l|}{Source Code} & \multicolumn{1}{c|}{0.29} & \multicolumn{1}{c|}{0.40} & \multicolumn{1}{c|}{0.43} & \multicolumn{1}{c|}{0.50} & \multicolumn{2}{c|}{N/A} & \multicolumn{2}{c|}{N/A} \\ \hline
\multicolumn{1}{|l|}{Dataset} & \multicolumn{1}{c|}{0.58} & \multicolumn{1}{c|}{0.77} & \multicolumn{1}{c|}{0.37} & \multicolumn{1}{c|}{0.71} & \multicolumn{1}{c|}{0.53} & \multicolumn{1}{c|}{0.79} & \multicolumn{1}{c|}{0.39} & \multicolumn{1}{c|}{0.88} \\ \hline
\multicolumn{1}{|l|}{Objective Task} & \multicolumn{2}{c|}{N/A} & \multicolumn{2}{c|}{N/A} & \multicolumn{1}{c|}{0.21} & \multicolumn{1}{c|}{0.59} & \multicolumn{1}{c|}{0.17} & \multicolumn{1}{c|}{0.85} \\ \hline
\multicolumn{1}{|l|}{Method Used} & \multicolumn{2}{c|}{N/A} & \multicolumn{2}{c|}{N/A} & \multicolumn{1}{c|}{0.17} & \multicolumn{1}{c|}{0.48} & \multicolumn{1}{c|}{0.13} & \multicolumn{1}{c|}{0.71} \\ \hline
\multicolumn{1}{|l|}{Computing Resources} & \multicolumn{1}{c|}{0.34} & \multicolumn{1}{c|}{0.53} & \multicolumn{2}{c|}{N/A} & \multicolumn{2}{c|}{N/A} & \multicolumn{2}{c|}{N/A} \\ \hline
\multicolumn{1}{|l|}{Language/Library} & \multicolumn{1}{c|}{0.10} & \multicolumn{1}{c|}{0.42} & \multicolumn{2}{c|}{N/A} & \multicolumn{2}{c|}{N/A} & \multicolumn{2}{c|}{N/A} \\ \hline
\multicolumn{1}{|l|}{Macro P \& R} & \multicolumn{1}{c|}{0.33} & \multicolumn{1}{c|}{0.53} & \multicolumn{1}{c|}{0.4} & \multicolumn{1}{c|}{0.61} & \multicolumn{1}{c|}{0.30} & \multicolumn{1}{c|}{0.62} & \multicolumn{1}{c|}{0.23} & \multicolumn{1}{c|}{0.81} \\ \hline
Macro F1 & \multicolumn{2}{c}{\textbf{0.41}} & \multicolumn{2}{c}{\textbf{0.48}} & \multicolumn{2}{c}{\textbf{0.40}} & \multicolumn{2}{c}{0.36} \\ \hline
\end{tabular}
\end{table*}

% \fi

% Researchers from Allen Institute For AI community have proposed an open-source document level dataset, SciREX~\cite{scirex}, for information extraction from scholarly articles. It comprises annotated salient entities like Task, Method, Dataset, and Metric. However, the \textsf{HPTLL} dataset focused on Hardware, Compute Time, Programming Language, and Libraries which has never been done before with on full-text scholarly articles. The SciREX model requires finer annotation to extract entities, and hence the SciREX dataset has more annotation. E2R requires coarser annotation to work, and hence HPTLL dataset has less annotation. Figure \ref{fig:e2r_vs_scirex_annotation} highlights the differences in the effort of an annotator to build the SciREX and the HPTLL(E2R) dataset.

\subsection{The ``Papers with Code'' dataset}

The ``Papers with Code'' dataset contains state-of-the-art papers for some research categories and highlights trends in research along with source code. We tested the extraction of source code links and dataset usage against a subset of the Papers with Code dataset. The results are presented in Table~\ref{tab:evaluation}.

For source code links, we randomly selected 980 papers from the nearly 15,000 papers with source code in the Papers with Code dataset. 
% Among these, \algo~ was able to extract at least one URL for 606 papers (61.83\%). Of these, 380 fully matched with ground truth URLs and 108 partially matched. Considering partial and full matches both as correct, the number of correct extractions is 488. The recall by paper is 0.498 and precision by paper is 0.805. Actual precision by number of extracted URLs is 0.43. 
For dataset usage, we were able to download 701 papers from 764 papers with such annotations. However, the Papers with Code dataset has limitations that hamper the completeness of information available for each paper.
Papers with Code maintains a list of datasets and lists papers underneath each one if the paper uses the dataset.
%Datasets used are presented in a top-down fashion, papers are listed starting from a dataset/task. 
This format often failed to capture all the dataset usage of a paper because either Papers with Code had no listing for a dataset used in a paper or papers simply are not listed under datasets which they did in fact use. For example, at least one paper in the corpus that uses four datasets (COCO, ImageNet, Kinetics, and Cityscapes) is listed under only one of those datasets in the Papers with Code database so taking Papers with Code as the ground truth shows only one dataset used in this paper: COCO. However, \algo~was able to pick up all four datasets from the paper. This gives our result a false impression of low precision.

\subsection{Comparison with SciREX dataset and model}
The SciREX system introduced a dataset with 438 papers and a model for information extraction from scientific articles. The dataset was built with entities from the ``Papers with Code'' dataset. The same limitations of the Papers with Code dataset apply here. It has four types of entities for each paper: method, metric, task and material. We evaluated our dataset, objective task, and method facets on a test set from the  SciREX dataset. To make an end-to-end comparison between SciREX model and \algo, we extracted ``salient entity clusters'' from SciREX. The ``salient entity clusters'' were then used in our evaluation process to generate evaluation metrics (section~\ref{sec:eval_metric}). The performance of SciREX model on SciREX dataset by our evaluation metrics are shown on the rightmost column of Table~\ref{tab:evaluation}. We found that SciREX usually churns out too many entity clusters which increases the recall but reduces precision and hampers the overall macro F1 for all facets in a dataset. In contrast, \algo~performed well over all the facets with an 11\% increase in the F1 value over SciREX.

% We do not need this metric
% following the same evaluation metric as described above, we computed at least one correct retrieval for 69.2\% of these papers. ``Total correct retrievals'' is 266.57, a 60.86\% success rate against the highest possible value of 438.

\section{Trends and key insights from scholarly literature}
\label{sec:experiment}
\algo~can be used to find trends and insights from a large number of papers. We attempted to answer several questions with the output of \algo~to showcase its utility and scalability. For source code links, dataset, method and task, the first four questions try to find existing trends by running our pipeline on the preprocessed arXiv CS papers. We also posed questions for usages of computing resource and language/library, specifically in the domain of AI. These questions are based on a subset of cs.AI papers from arXiv. 
\newline

\begin{figure*}[t]
    \centering
    \begin{subfigure}[t]{0.3\textwidth}
        \includegraphics[trim={0.2cm 0.25cm 0.5cm 1.4cm},clip,width=\textwidth]{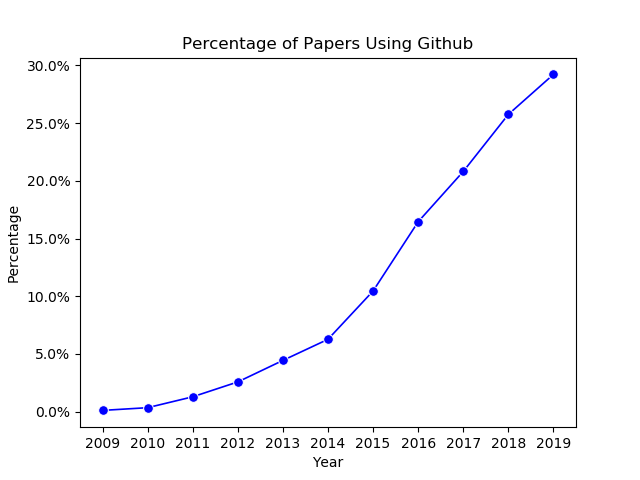}
        \caption{Papers with source code hosted \\on GitHub.}
        \label{fig:github_trend}
    \end{subfigure}
    \begin{subfigure}[t]{0.3\textwidth}
        \includegraphics[trim={0.2cm 0.25cm 0.5cm 1.4cm},clip,width=\textwidth]{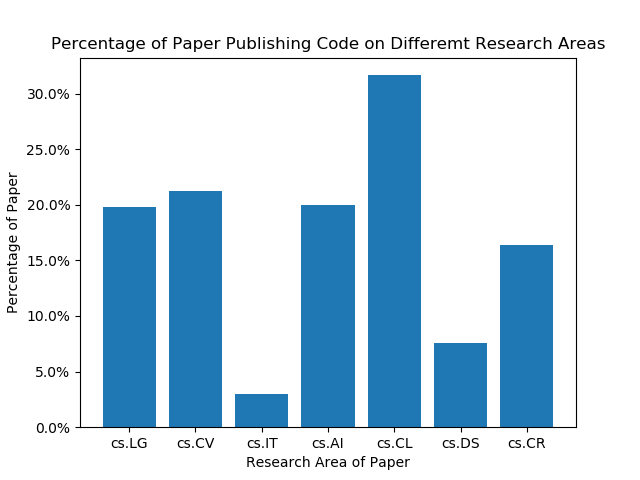}
        \caption{Percentage of papers containing source code links for various research areas.}
        \label{fig:csTag}
    \end{subfigure}
    \begin{subfigure}[t]{0.3\textwidth}
         \includegraphics[trim={0.25cm 0.25cm 0.5cm 1.4cm},clip,width=\textwidth]{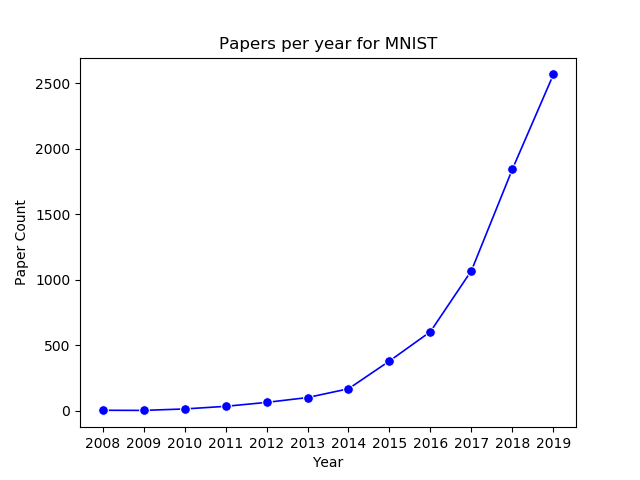}
        \caption{Papers per year using MNIST.}
        \label{fig:mnist_trend}
    \end{subfigure}
    
    \begin{subfigure}[b]{0.3\textwidth}
        \includegraphics[trim={0.25cm 0.25cm 0.5cm 1.4cm},clip,width=\textwidth]{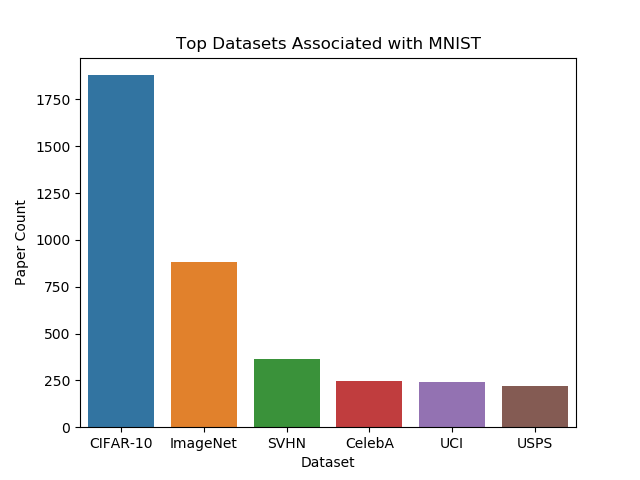}
        \caption{Top datasets associated with the MNIST dataset.}
        \label{fig:data_mnist}
    \end{subfigure}
    \begin{subfigure}[b]{0.3\textwidth}
        \includegraphics[trim={0.25cm 0.25cm 0.5cm 1.4cm},clip,width=\textwidth]{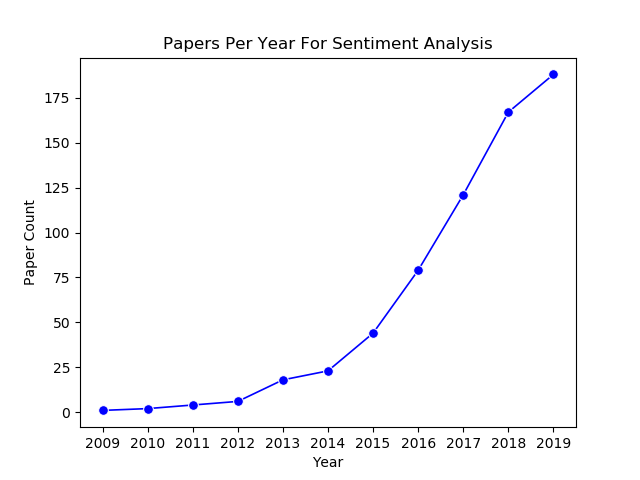}
        \caption{Count of papers on sentiment analysis over time.}
        \label{fig:sentiment_trend}
    \end{subfigure}
    \begin{subfigure}[b]{0.3\textwidth}
        \includegraphics[trim={0.25cm 0.25cm 0.5cm 1.4cm},clip,width=\textwidth]{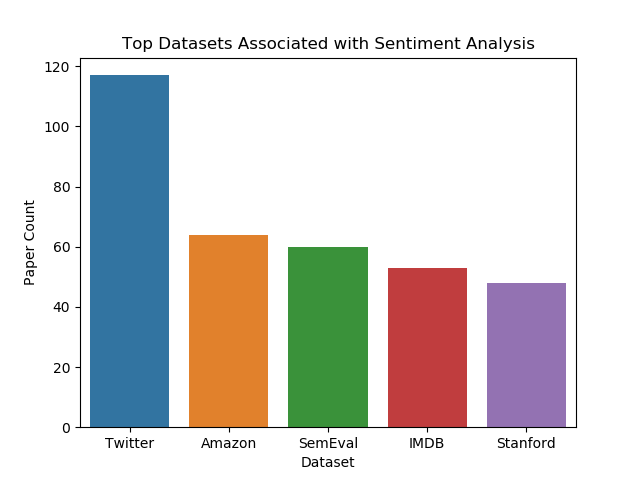}
        \caption{Top datasets for sentiment analysis.}
        \label{fig:data_sentiment}
    \end{subfigure}
    
    \begin{subfigure}[b]{0.3\textwidth}
        \includegraphics[trim={0.25cm 0.25cm 0.5cm 1.4cm},clip,width=\textwidth]{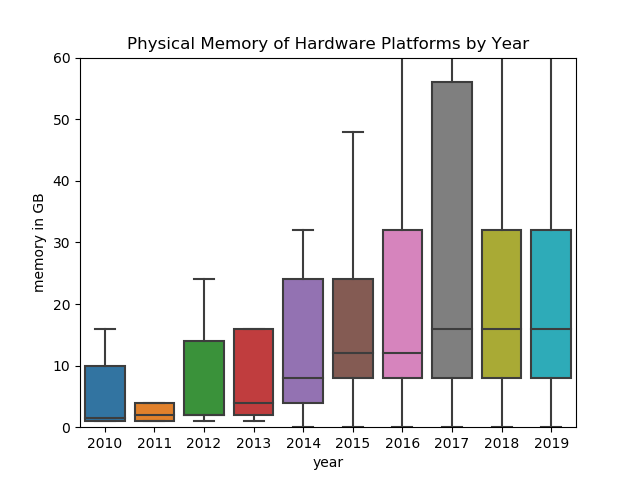}
        \caption{Hardware memory by year.}
        \label{fig:memory}
    \end{subfigure}
    \begin{subfigure}[b]{0.3\textwidth}
        \includegraphics[trim={0.25cm 0.25cm 0.5cm 1.4cm},clip,width=\textwidth]{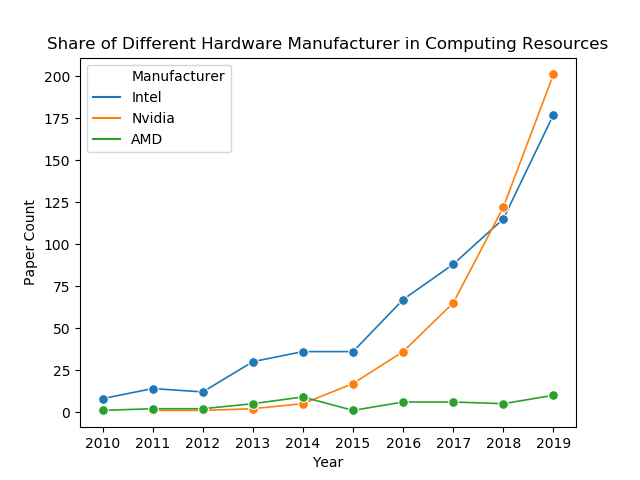}
        \caption{Popularity of hardware manufacturers by paper count.}
        \label{fig:intel_nvidia}
    \end{subfigure}
    \begin{subfigure}[b]{0.3\textwidth}
        \includegraphics[trim={0.25cm 0.25cm 0.5cm 1.4cm},clip,width=\textwidth]{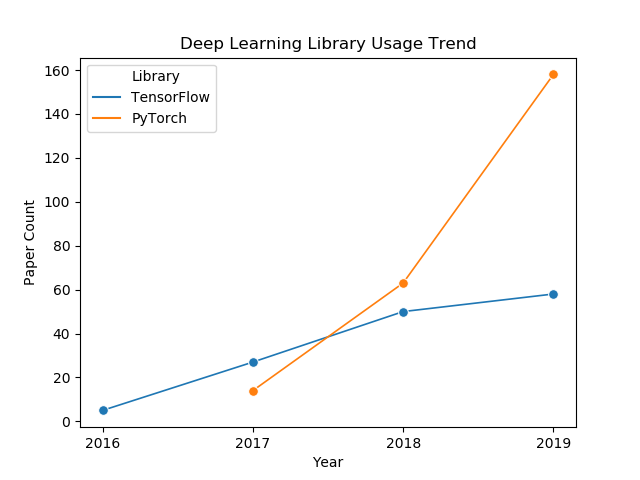}
        \caption{Deep learning library usage (TensorFlow vs PyTorch)}
        \label{fig:tensorflow_pytorch}
    \end{subfigure}
    
    \begin{subfigure}[b]{0.3\textwidth}
        \includegraphics[trim={0.25cm 0.25cm 0.5cm 1.4cm},clip,width=\textwidth]{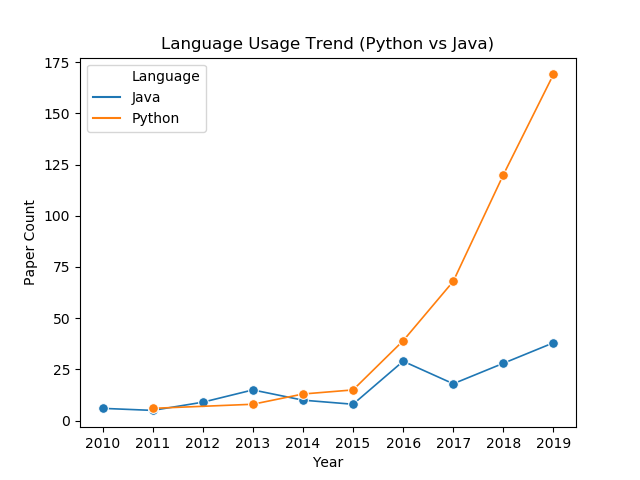}
        \caption{Programming language usage (Java vs Python)}
        \label{fig:python_java}
    \end{subfigure}
    %\begin{subfigure}[b]{0.32\textwidth}
    %\end{subfigure}
    \caption{Evaluation of \algo}
\end{figure*}

\textbf{Q1. What is the usage of GitHub as choice of platform to share code?}
We looked into the output of the source code link to find out how many of the papers are using GitHub as platform to share source code. We found that the percentage of papers sharing source code links with GitHub is increasing each year. The trend line is presented in Figure \ref{fig:github_trend}.
\newline

%\begin{figure}[htp!]
%  \centering
%  \includegraphics[trim={0.2cm 0.25cm 0.5cm 1.4cm},clip,width=0.9\linewidth]{Images/github.png}
%  \caption{Papers with source code hosted on GitHub.}
%  \label{fig:github_trend}
%\end{figure}

\textbf{Q2. Does one area of research share source code links more often than others?}
% We again considered the output of the source code links to answer this question. 
To identify and designate a paper to a research area, we used the metadata available from arXiv, which tags each paper with several tags indicating the research areas. We calculated the total number of papers for each of the 40 computer science subcategories and found the percentage of papers publishing a source code link. ``Computation and language'' has the highest percentage with 31.64\%, followed by ``computer vision'' with 21.26\%, ``artificial intelligence'' with 19.99\%, and ``machine learning'' with 19.79\%. Figure \ref{fig:csTag} summarizes the findings.
\newline

%\begin{figure}[htp!]
%  \centering
%  \includegraphics[trim={0.2cm 0.25cm 0.5cm 1.4cm},clip,width=0.9\linewidth]{Images/csTag.png}
%  \caption{Percentage of papers that contain source code links for various research areas.}
%  \label{fig:csTag}
%\end{figure}

\textbf{Q3. What is the most used dataset? What are the top objective tasks, methods and other datasets used along with it?} 
% To answer this question we looked into the extracted dataset entities.
We found out that MNIST is the most used dataset in our full arXiv corpus. In fact, papers that use MNIST have increased over time as shown in Figure~\ref{fig:mnist_trend}.

%\begin{figure}[htp!]
%  \centering
%  \includegraphics[trim={0.25cm 0.25cm 0.5cm 1.4cm},clip,width=0.9\linewidth]{Images/trend_MNIST.png}
%  \caption{Papers per year using MNIST.}
%  \label{fig:mnist_trend}
%\end{figure}

The top used datasets after MNIST are CIFAR and ImageNet. We present other top datasets used in these papers in Figure~\ref{fig:data_mnist}. The top tasks for MNIST papers are computer vision, image classification, and object detection. Similarly we found that the top methods are CNN, deep neural network, and GAN. 
% We also demonstrated these task and method with two corresponding word clouds in Figure \ref{fig:task_mnist} and \ref{fig:method_mnist}.
\newline

%\begin{figure}[htp!]
  %\centering
  %\includegraphics[trim={0.25cm 0.25cm 0.5cm 1.4cm},clip,width=0.9\linewidth]{Images/dataset_bar_chart_MNIST.png}
  %\caption{Top datasets associated with the MNIST dataset.}
%  \label{fig:data_mnist}
%\end{figure}

%%%%%%%%%%%%%%%%%%%%
% \begin{figure}
%      \centering
%          \begin{subfigure}[b]{0.495\linewidth}
%              \centering
%              \includegraphics[trim= 1.5cm 1.5cm 1.5cm 1.5cm, clip,width=\textwidth]{Images/taskCloud_MNIST.png}
%              \caption{}
%              \label{fig:task_mnist}
%          \end{subfigure}
%     \hfill
%          \begin{subfigure}[b]{0.495\linewidth}
%              \centering
%              \includegraphics[trim= 1.5cm 1.5cm 1.5cm 1.5cm, clip,width=\textwidth]{Images/methodCloud_MNIST.png}
%              \caption{}
%              \label{fig:method_mnist}
%          \end{subfigure}
%     \caption{Word cloud associated with MNIST dataset for (a) top tasks, and (b) top methods.}
%     \label{fig:three graphs}
% \end{figure}

\textbf{Q4. What are the research trends in sentiment analysis? What are the top datasets used for this topic?
}
We created a trend line with the number of papers working on sentiment analysis, a classic research problem in the NLP community, over the years in Figure~ \ref{fig:sentiment_trend}. 
%We used the objective task output of \algo~ and created a trend line with the number of papers working on it 
Among the papers, Twitter is the most used dataset, followed by Amazon, SemEval, IMDB, and Stanford. We presented these findings in Figure \ref{fig:data_sentiment}.
\newline

%\begin{figure}[htp!]
%  \centering
%  \includegraphics[trim={0.25cm 0.25cm 0.5cm 1.4cm},clip,width=0.9\linewidth]{Images/sentiment.png}
%  \caption{Count of papers on sentiment analysis over time.}
%  \label{fig:sentiment_trend}
%\end{figure}

%\begin{figure}[htp!]
%  \centering
%  \includegraphics[trim={0.25cm 0.25cm 0.5cm 1.4cm},clip,width=0.9\linewidth]{Images/dataset_bar_chart_sentiment.png}
%  \caption{Top datasets for sentiment analysis.}
%  \label{fig:data_sentiment}
%\end{figure}

\textbf{Q5. How much memory do researchers use in AI research?}
\algo`s computing resource output contains information on the amount of physical memory (RAM) and GPU memory used for simulations. We did not differentiate between the GPU memory and physical memory for this use case. So, this may be the maximum memory of the hardware platform used by each paper rather than the actual used memory for the systems. We found out that the median physical memory has been increasing almost each year until 2017. 2019 also saw same median and same third quartile values for memory used throughout the community which is 16GB and 32GB respectively. A boxplot is shown in Figure~\ref{fig:memory}.
\newline

%\begin{figure}[htp!]
%  \centering
%  \includegraphics[trim={0.25cm 0.25cm 0.5cm 1.4cm},clip,width=0.9\linewidth]{Images/memory.png}
%  \caption{Hardware memory by year.}
%  \label{fig:memory}
%\end{figure}

\textbf{Q6. What is the market share of different hardware manufacturers in the AI research community?} 
We looked in our computing resource output specifically for \textsf{Intel}, \textsf{Nvidia}, and \textsf{AMD}. We excluded papers which mentioned two of the brands and considered papers using only one of these brands. We found that historically Intel has been the strongest in market share in CS research community. However, in 2018, Nvidia surpassed Intel as the most used hardware platform and the upward trend is continuing, possibly due to an increase in deep learning frameworks which rely on high GPU usage. On the other hand, AMD does not hold any significant share of the research platform. The trend is shown in Figure~\ref{fig:intel_nvidia}.
\newline

%\begin{figure}[htp!]
%  \centering
%  \includegraphics[trim={0.25cm 0.25cm 0.5cm 1.4cm},clip,width=0.9\linewidth]{Images/intelvnvidia.png}
%  \caption{Popularity of hardware manufacturers by paper count.}
%  \label{fig:intel_nvidia}
%\end{figure}

\textbf{Q7. What is the relative usage of popular deep learning frameworks in AI research?}
We compared the usage of two most popular deep learning frameworks: TensorFlow and PyTorch. 
%We used the output from language/library to find out the trend over the years. 
The trend as paper count over time  is presented in Figure~\ref{fig:tensorflow_pytorch}. Based on our findings, TensorFlow was more popular than PyTorch before 2018. Since then PyTorch has surpassed TensorFlow with a higher growth rate and has continued its climb over the following year.
\newline

%\begin{figure}[htp!]
%  \centering
%  \includegraphics[trim={0.25cm 0.25cm 0.5cm 1.4cm},clip,width=0.9\linewidth]{Images/tensorflowvpytorch.png}
%  \caption{Deep learning library usage (TensorFlow vs PyTorch)}
%  \label{fig:tensorflow_pytorch}
%\end{figure}

\textbf{Q8. Which programming language is the most popular in AI research?}
We compare the number of papers for which we extracted Java or Python for the language/library facet in Figure~\ref{fig:python_java}.
%make a comparison with the usage by the research community. 
% We used the output from language/library to map the trend for this two programming languages. 
Java was the more popular language until 2013. Python narrowly beat Java in 2014 and 2015 from which point Python started growing with a sharp upward trend. In 2019, four times as many papers mentioned Python as Java.

%\begin{figure}[htp!]
%  \centering
%  \includegraphics[trim={0.25cm 0.25cm 0.5cm 1.4cm},clip,width=0.9\linewidth]{Images/python_java.png}
%  \caption{Programming language usage (Java vs Python)}
%  \label{fig:python_java}
%\end{figure}

\section{Lessons Learnt \& Future Research}
\label{sec:discussion}
Here we present lessons learnt during our attempt at entity extraction from full-text documents. Furthermore, the comparison with multiple dataset and state-of-the-art information extraction systems also pointed out some important aspects of existing systems in contrast to our approach. We discuss these issues below.

\subsection{Lack of ground truth}
The foremost difficulty involved in this type of full-text extraction is the lack of an established ground truth dataset which hampers both development and the evaluation of such a project. To the best of our knowledge, the SciREX and Papers with Code datasets are the only datasets in this area. As described, the Papers with Code dataset is not complete so could give a false impression of low precision. SciREX suffers from the same issues as it was built with Papers with Code entities. Moreover, three of our facets have not been well studied in the literature. Because of these limitations, we developed our own training data generator to train classifiers with a weakly supervised learning paradigm. 

% include recommendation for future research in different kinda text

\subsection{Issue with automated evaluation}
% Need to show example and verification (and/or stats for this, working)
We discovered that the ground truths in the Paper with Code dataset are generalized versions of the entities and these are traditionally decided by community members. The entities may not appear exactly as they do in the paper. SciREX was built with these same entities. For SciREX, we observed that 66.72\% task entities, 61.16\% method entities, and 84.12\% material entities do not appear as the ground truth in the paper. However, \algo~can only extract what is in the text of the paper. Consequently, for some papers, the automated evaluation system failed to correctly capture the comparison between ground truth and \algo-extracted entities. We partially solved this issue by adopting a fuzzy string matching based evaluation process (see Section~\ref{sec:eval_metric}) and cleaning the ground truth values.

% example of failure on partial matching: 0910a4c470a410fac446f4026f7c8ef512ae7427, 0e6b990f41fc0e7fdd0662f88b5ad024375bd77b
% dataset named as semeval_task: 1512.01100, "0b0dc14b8a8dcccbfc62a38355fff2f6a361e9d2", Paper did not mention semeval at all, only mention "benchmark dataset on Twitter". Our algo picked "Twitter"
% similarly, 1510cf4b8abea80b9f352325ca4c132887de21a0, gold: qasent and wikiqa, Our algo picks up perfectly two used dataset, in pwc, qasent is not even connected with any dataset and wikiqa was not used in this paper at all

\subsection{Full Text vs Abstract vs Specific Sections}
The choice of the input for each facet should be selected following the usual structure of scholarly articles. We ingested full text to create a representation which is structured by sections.
Firstly, using full text is necessary for the extraction of some facets as they are generally only mentioned in the full text, especially source code, dataset usage, hardware related entities. Secondly, working with full text in machine learning models is sometimes too computation-heavy. Moreover, some sections related to methodological or mathematical components generally introduce noise for some facets. 
For objective tasks and methods, researchers tend to summarize and mention the main objective tasks and goal in the abstract, introduction and conclusion sections. With a sample set of papers, we checked the extraction rate of~\algo~when processing abstracts alone vs processing abstracts, introductions, and conclusions and found that the inclusion of introductions and conclusions improves the extraction rate by 26\%. We therefore struck a balance between these practical findings and only used the sections that were most likely to contain the specific information we seek for each facet.

% please go through why did you choose PDF, add some statistics later
\subsection{Full Text Ingestion}
The most common digital object format is PDF but it does not save any structural information along with its graphical representation. Even using a state-of-the-art PDF ingestion system (Grobid), we found instances where  footnotes were not extracted properly. References and URLs were also often incorrectly extracted. Improvements to this part of the ingestion system, may also improve \algo's performance.

\subsection{Entity Variations and Sub-Entities}
% A particular challenge is the variation of the same entity names. 
\algo~often picks up different variations of the same entity. 
% Most of the entity can be mentioned in many different ways. 
For instance, datasets can be mentioned in abbreviated form or full form or  may have designated subsets. Objective tasks can be phrased in different ways. Moreover, each entity can have multiple sub-entities. We solved this issue by clustering such lists of entities and considering them a single entity. However, we did not particularly identify the sub-entities and main entity those belong to. Ideally, we would like to map each variation of an entity to a standard entity list but making this entity list remains as a future task. 

\subsection{Division of Tasks and Computational Load}
Instead of running a single pipeline to churn out all the facets at once, we divided our pipeline into manageable modules. The sentence classifier part detects salient sentences first and those sentences are fed into an NER module to extract word level entities. Moreover, we further divided our pipeline into separate parallel executable modules for each facet. This keeps our pipeline simpler and annotation effort can be precise for each facet as required. However, by doing this and not merging the facets, we lose relationships between them. Only a few of our facets have a obvious relationship to each other such as computing resources and language/library can be a group which sometimes have a ``feature-of'' or ``used'' relationship within the entities. Similarly dataset, objective task and method could be combined in a relationship. We leave this task for future expansion.

\subsection{Knowledge Graph and Application}
We posit that our extracted facets can be used to enrich existing knowledge graphs and create new layers of information. Here we will show one use case of how researchers are using such knowledge graphs and where our developed \algo~fits into equation. 

%\begin{figure}
%
%\subfloat[Overview of Map of Science]{\includegraphics[trim={0cm 0cm 0cm 0.07cm},clip,height=6cm, width=0.6\columnwidth]{Images/map_of_sci.jpg}}

%\subfloat[Papers Published per Year by Topic, 2010–2020]{ %\includegraphics[clip,width=0.9\columnwidth]{Images/gelles.jpg}}

%\caption{Use of a knowledge graph, taken from %\cite{gelles_measuring_ai}}
%\label{fig:gelles}
%\end{figure}

A recent article from a policy research organization focused on measuring the development of several AI-related topics: ``re-identification'', ``speaker recognition'', and ``image synthesis''~\cite{gelles_measuring_ai}. The authors singled out the CSET Map of Science\cite{aI_funding_research} clusters for each topic using seed papers and carried out manual analysis to map out progress in each of these topics.
%A view of the number of papers found for first two topics is shown in Figure~\ref{fig:gelles}. 
However, the authors noted that such a clustering tool cannot fully encompass the subject area and struggled to find a single representative cluster for ``image synthensis''. Moreover, to establish the quality of a research topic, performance metrics (i.e., dataset used) need to be tracked which was done manually in the paper. By contrast, \algo~can extract objective task, methods, and datasets used. The first two facets can be used to filter out a list of papers with that particular objective task (and/or method). \algo~can also produce lists of papers that use a specific dataset. The authors also recommended the establishment of a continuous analysis system to keep track of research development. Using an automated system like \algo~to find out the interesting facets would enable us toward that goal.

After incorporating the extracted entities of the papers into such a knowledge graph, the next step will be to translate the extracted raw entities back to a standardized list of generic entities so that a hierarchical taxonomy of the facets and sub-facets can be created within the knowledge graph. But there is no established taxonomy available for standardizing different scientific facets, sub-facets and their variations, i.e., objective tasks, methods and application areas. Most researchers adopt ad-hoc taxonomies suited to their specific research goal. Thus this issue remains an open research problem for future expansion.

\section{Conclusion}
\label{sec:conclusion}
This work presents an information extraction pipeline, \algo, developed for extracting six scientific facets from full-text scholarly research articles: link to source code, dataset used, objective task, method, computing resources and language/library dependency. We demonstrated how the results from \algo~can be helpful in discovering research trends and emerging technologies from a large scale dataset. We evaluated \algo~on multiple datasets and highlighted the shortcomings in existing datasets.
% also write if you provide any insights for future development
In addition to establishing trends, extracted entities from large scale scholarly datasets can be used to build knowledge graphs which have tremendous promise for the research community. We discussed how \algo~can complement such knowledge graphs with a use case.
%For e.g., we can quickly find out the research papers for a particular tasks, method or find out the source code and the dataset used for the paper.
This research for full-text entity extraction serves as a stepping stone towards automated systems for extracting entities and updating such knowledge graphs without human effort.

% CSET will be in the authors
\begin{acks}
 We are grateful to Catherine Aiken, Chengzhen Bian, Daniel Chou and Jennifer Melot for insightful comments and helpful discussions that shaped the research effort. 
This work was performed under a subgrant agreement with the Center for Security and Emerging Technology (CSET) at Georgetown University, subgrant number AWD7773402-GR206518. %\textbf{Disclaimer:} The views and conclusions contained in this document are those of the authors and should not be interpreted as representing the official policies, either expressed or implied, of CSET, Virginia Tech or the US Government.
\end{acks}


\bibliographystyle{ACM-Reference-Format}
\bibliography{sample-base}

\appendix
\section{Model Details}
For sentence classifiers, we initialized the model with SciBERT base \cite{beltagy2019scibert}. We isolated the candidate sentences from our training data generation script and randomly sampled twice as many incorrect sentences from the papers. The train, dev and test split is (85\%:10\%:5\%). For dataset and source code links, We trained the models for 3 epochs using AdamW optimizer with learning\_rate 2e-5 and seed being 1. The batch size was 8. For computing resources and language/library, we trained the models for 10 epochs with batch size 32. For the dataset named entity recognizer, we tokenized the candidate sentences from our weakly supervised learning with Spacy \cite{spacy}, tagged the tokens in BILUO format and trained a transformer model, initialized with SciBERT for 3 epochs with batch size 16. The maximum sequence length is 256 for both cases. For computing resource and language/library, we trained the model with CORLL dataset with batch size 32.

To evaluate, we trained the SciREX model ourselves following the guideline in the paper. We trained the main model for 20 epochs and the coreference model for 10 epochs, using 10 and 4 as patience value respectively. To ensure an even comparison, we identified the salient entity clusters from the dataset and feed them into our own evaluation metric which used a fuzzy string matching. All of our training and prediction is using a NVIDIA Tesla P100 GPU with 16GB memory.

\section{Annotated Dataset Statistics} \label{app:dataset_statistics}
We annotated a dataset, specifically for the evaluation of our computing resource and language/library facets. The distribution of different facets inside this dataset are presented below.

\begin{table}[htpt!]
\centering
\caption{Facets and their distribution in the annotated ground truth for evaluation}
\begin{tabular}{|c|c|}
\hline
\textbf{Facet Type} & \textbf{Percentage} \\ \hline
Datasets & 50.34\% \\ \hline
Source Code & 6.89\% \\ \hline
Computing Resources &  28.97\% \\ \hline
Language/Library & 40.69\% \\ \hline
\end{tabular}%
\label{tab:annotation}
\end{table}

\section{Syntactic Pattern Creation} \label{app:patterns}
For developing the syntactic patterns, we focused on four major CS areas in which discussion of these facets is more prevalent: Computer Vision, Machine Learning, Hardware Architecture, and Artificial Intelligence. Each of these areas can be identified via arXiv metadata tags. To design the patterns, we selected 80 documents for each of the facets (20 each from the four aforementioned areas) via some sample seed words related to that facet. As a next step, we isolated the sentences containing these seed words and identified their syntactic properties and patterns. The patterns and templates for each of the facets created by this process are the building blocks of our syntactic pattern based extraction algorithms. For source code links, only sentences sufficed as ground truth. For dataset, computing resources and language/library, we required both sentence level and entity level ground truth to train our models.

\begin{table}[htpt!]
\centering
\caption{Syntactic Patterns for Source Code Extraction}
\vspace{-1em}
\label{tab:pattern_source}
\begin{tabular}{|p{0.25\columnwidth}|p{0.65\columnwidth}|}

\hline
\textbf{Pattern }  & \textbf{Example }                                                                      \\ \hline \hline
\textbf{(subj)} (*) (root) & it we model implementation source code supplementary material                 \\ \hline
\textbf{(root)}                                                             & is are find release                                                           \\ \hline
(root) (*) \textbf{(obj)}  & Github website open-source implementation project page supplementary material \\ \hline
\textbf{(adj/adv)}                                                        & publicly available online opensource open-source supplementary                \\ \hline
\end{tabular}
\end{table}

\begin{table}[htpt!]
\centering
\caption{Syntactic Patterns for Dataset Extraction}
\vspace{-1em}
\label{tab:pattern_data}
\begin{tabular}{|p{0.3\columnwidth}|p{0.6\columnwidth}|}

\hline
\textbf{Pattern }                          & \textbf{Example }                                                               \\ \hline \hline
\textbf{(subj)} (*) (root)                & performance paper we dataset experiment                                                                                                       \\ \hline
\textbf{(root) }                          & make utilize adopt create construct include consist perform introduce contain feed is use implement evaluate release focus conduct constitute \\ \hline
(root) (*) \textbf{(obj)}                 & database dataset github website repository online collection benchmark numerical study                                                        \\ \hline
\textbf{(adj/adv) }                       & publicly available online large-scale constructed synthetic dataset popular constructed                                                       \\ \hline
(root) (*) (which/that) \textbf{(verb)}   & generate provide utilize adopt create construct include consist introduce contain feed use release                                            \\ \hline
\textbf{(root)} (*) (number)              & include consist contain constitute compose comprise                                                                                           \\ \hline
(root) (*) \textbf{(adj clause)} (number) & compose consist comprise                                                                                                                      \\ \hline
\end{tabular}
\end{table}

\section{Evaluation Metrics} \label{app:eval}
For links to source code, we considered an extraction as correct if at least the first part of the path (excepting the network location) matches the ground truth. For example, we consider the following case as correct.

ground truth:``github.com/pwc/pwc-data''

extraction: ``github.com/pwc''

\noindent But the following extraction is not correct:
%We consider these type of partial matched and fully matched both as correct.

ground truth:``github.com/pwc/pwc-data''

extraction: ``github.com''
% \include{appendixB}
			%\lipsum[2]

\end{document}